\documentclass[twocolumn]{aastex631}

\usepackage[T1]{fontenc}

\DeclareRobustCommand{\VAN}[3]{#2}
\let\VANthebibliography\thebibliography
\def\thebibliography{\DeclareRobustCommand{\VAN}[3]{##3}\VANthebibliography}

\usepackage{amsmath}
\usepackage{aas_macros}
\usepackage{natbib}
\usepackage{xspace}
\usepackage[varg]{txfonts}
\usepackage{graphicx}
\usepackage{nicefrac}
\usepackage{tablefootnote}
\usepackage{multirow}
\usepackage{soul}
\usepackage{float}
\usepackage[T1]{fontenc}
\usepackage{lmodern}
\usepackage{multirow}
\usepackage{ulem}
\usepackage{soul}

\newcommand{\msolar}{M$_{\odot}$}

\begin{document}

$^{2}$ofox@stsci.edu.\\


\shorttitle{SN 2013ge Companion Star}
\shortauthors{O. D. Fox et al.}

\title{The Candidate Progenitor Companion Star of the Type Ib/c SN 2013ge}

\author[0000-0003-2238-1572]{Ori D. Fox}
\affiliation{Space Telescope Science Institute, 3700 San Martin Drive, Baltimore, MD 21218, USA}

\author[0000-0001-9038-9950]{Schuyler D. Van Dyk}
\affiliation{Caltech/IPAC, Mailcode 100-22, Pasadena, CA 91125, USA}

\author[0000-0002-7502-0597]{Benjamin F. Williams}
\affiliation{University of Washington, Seattle, WA  98195, USA}

\author[0000-0001-7081-0082]{Maria Drout}
\affiliation{David A. Dunlap Department of Astronomy and Astrophysics, University of Toronto \\ 50 St. George Street, Toronto, Ontario, M5S 3H4 Canada}
\affiliation{Observatories of the Carnegie Institution for Science, 813 Santa Barbara St., Pasadena, CA 91101, USA}

\author[0000-0002-7464-498X]{Emmanouil Zapartas}
\affiliation{Département d’Astronomie, Université de Genève, Chemin Pegasi 51, CH-1290 Versoix, Switzerland}

\author[0000-0001-5510-2424]{Nathan Smith}
\affiliation{Steward Observatory, University of Arizona
933 N. Cherry Ave., Tucson, AZ 85721, USA}

\author[0000-0002-0763-3885]{Dan Milisavljevic}
\affiliation{Department of Physics and Astronomy, Purdue University, 525 Northwestern Avenue, West Lafayette, IN 47907-2036, USA}
\affiliation{Integrative Data Science Initiative, Purdue University, West Lafayette, IN 47907, USA}

\author[0000-0003-0123-0062]{Jennifer E. Andrews}
\affiliation{Gemini Observatory/NSF’s NOIRLab, 670 N. A’ohoku Place, Hilo, HI 96720, USA}

\author[0000-0002-4924-444X]{K. Azalee Bostroem}
\affiliation{DIRAC Institute, Department of Astronomy, University of Washington, 3910 15th Avenue NE, Seattle, WA 98195, USA)}

\author[0000-0003-3460-0103]{Alexei V. Filippenko}
\affiliation{Department of Astronomy, University of California, Berkeley, CA 94720-3411, USA}

\author[0000-0001-6395-6702]{Sebastian Gomez}
\affiliation{Space Telescope Science Institute, 3700 San Martin Drive, Baltimore, MD 21218, USA}

\author[0000-0003-3142-997X]{Patrick L. Kelly}
\affiliation{Minnesota Institute for Astrophysics, 116 Church St. SE, Minneapolis, MN 55455, USA}

\author[0000-0001-9336-2825]{S. E. de Mink}
\affiliation{Max Planck Institute for Astrophysics, Karl-Schwarzschild-Str. 1, 85748 Garching, Germany}
\affiliation{Anton Pannekoek Institute for Astronomy, University of Amsterdam, Science Park 904, 1098XH Amsterdam, The Netherlands}

\author[0000-0002-2361-7201
]{Justin Pierel}
\affiliation{Space Telescope Science Institute, 3700 San Martin Drive, Baltimore, MD 21218, USA.}

\author[0000-0002-4410-5387 ]{Armin Rest}
\affiliation{Space Telescope Science Institute, 3700 San Martin Drive, Baltimore, MD 21218, USA.}
\affiliation{Department of Physics and Astronomy, The Johns Hopkins University, Baltimore, MD 21218, USA}

\author[0000-0003-4501-8100]{Stuart Ryder}
\affiliation{School of Mathematical and Physical Sciences, Macquarie University, NSW 2109, Australia}
\affiliation{Astronomy, Astrophysics and Astrophotonics Research Centre, Macquarie University, Sydney, NSW 2109, Australia}

\author{Niharika Sravan}
\affiliation{Division of Physics, Mathematics, and Astronomy, California Institute of Technology, Pasadena, CA 91125, USA}

\author{Lou Strolger}
\affiliation{Space Telescope Science Institute, 3700 San Martin Drive, Baltimore, MD 21218, USA.}

\author[0000-0001-5233-6989]{Qinan Wang}
\affiliation{Department of Physics and Astronomy, The Johns Hopkins University, Baltimore, MD 21218, USA}

\author[00000-0002-4471-9960]{Kathryn E.\ Weil}
\affil{Department of Physics and Astronomy, Purdue University, 525 Northwestern Avenue, West Lafayette, IN 47907, USA}

\begin{abstract} 

This {\it Letter} presents the detection of a source at the position of the Type Ib/c supernova (SN) 2013ge more than four years after the radioactive component is expected to have faded. This source could mark the first post-SN direct detection of a surviving companion to a stripped-envelope Type Ib/c explosion. We test this hypothesis and find the shape of the source's spectral energy distribution is most consistent with that of a B5~I supergiant. While binary models tend to predict OB-type stars for stripped-envelope companions, the location of the source on a color-magnitude diagram (CMD) places it redward of its more likely position on the main sequence (MS). The source may be temporarily out of thermal equilibrium, or a cool and inflated non-MS companion, which is similar to the suggested companion of Type Ib SN 2019yvr that was constrained from pre-SN imaging. We also consider other possible physical scenarios for the source, including a fading SN, circumstellar shock interaction, line of site coincidence, and an unresolved host star cluster, all of which will require future observations to more definitively rule out. Ultimately, the fraction of surviving companions (``binary fraction'') will provide necessary constraints on binary evolution models and the underlying physics.
\end{abstract}


\section{Introduction}\label{sec:intro}

 Core-collapse supernovae (CCSNe) arise from the deaths of massive stars ($M_{\rm ZAMS} \gtrsim 8$\,\msolar). Stripped-envelope explosions --- SNe~IIb, Ib, and Ic --- refer to the subset of CCSNe with progenitors that have lost all, or almost all, of their outer hydrogen and helium envelopes in pre-SN mass loss.  Traditional single-star evolution models explain this mass loss with line-driven winds \citep[e.g.,][]{heger03}, where progenitor mass and metallicity correlate with the degree of mass lost.  

 New evidence over the past 15 years has begun to shift this paradigm.  For example, \citet{smartt09} highlights a lack of massive, single-star progenitors to SNe~Ib/c detected in pre-explosion imaging. \citet{smith11} find that an initial mass function (IMF) of solely single stars cannot adequately reproduce the observed fractions of stripped-envelope SNe. \citet{drout11} show that SNe~Ib/c have relatively low ejecta masses; stellar winds are generally too weak to remove the entire outer envelope on their own \citep{smith14}, except perhaps in the most massive (and rare) stars. \citet{sana12} report that $>30$\% of massive stars form in some sort of {\it interacting} binary system. 
 
Binary-star physics is important for our understanding of massive-star evolution and many areas of astrophysics, from galactic chemical evolution to gravitational-wave detection \citep{belczynski16}. While massive binary star models for SN progenitors have existed for decades \citep[e.g.,][]{podsiadlowski92}, the detailed physics (e.g., winds, mass exchange, rotation) remain unconstrained \citep[e.g.,][]{zapartas17b,eldridge17}. A comprehensive, statistically complete sample of direct companion observations is necessary to measure the binary fraction, stellar type, and mass distribution, but such observations remain sparse.
 
 The search for a surviving binary companion star to a stripped-envelope SN is demanding, requiring high-resolution and subpixel astrometric precision. Furthermore, optical searches are likely insufficient, as potential surviving companions may be hot, blue main-sequence (MS) stars with a mass distribution peaking at $\sim9$--10\,\msolar, which typically correspond to O-type and B-type stars peaking at ultraviolet (UV) wavelengths $<2000\,\AA$ \citep[e.g.,][]{zapartas17b}. To date, only a handful of candidate companion direct detections exist: the Type IIb 1993J \citep{vandyk02, maund04, fox14}, the Type IIb 2001ig \citep{ryder18}, the Type IIb 2011dh \citep[][]{bersten12, benvenuto13, folatelli14}, and the Type Ibn SN 2006jc \citep{maund16,sun20}. The nature of some, if not all, of these sources is still ambiguous. In any case, no post-explosion detections of a binary companion exist yet for fully stripped SNe~Ib/c, although pre-explosion spectral energy distributions (SEDs) suggest the possible presence of a companion source in the case of SN~Ib iPTF13bvn \citep{cao13,eldridge16,folatelli16} and SN~Ib 2019yvr \citep{kilpatrick21,sun21}.
 
SN 2013ge is a Type Ic (or possibly a Ib) SN discovered on 2013 Nov. 8.796 2013 (UT dates are used throughout this paper) in NGC~3287 \citep{nakano13}.  Multiwavelength (radio to X-ray) observations were obtained from $-13$ to +457\,days relative to maximum light, including a series of optical spectra and {\em Swift} UV–optical photometry beginning 2–-4\,days post-detection. \citet{drout16} presented a detailed analysis of these data.  The properties of the SN 2013ge light curves are within the distribution observed for SNe~Ib/c, but several unique features stand out. The spectrum is predominantly that of a fully stripped SN~Ic, but \citet{drout16} give it a Type Ib/c classification owing to evidence for weak He features at early times. \citet{drout16} derive a pseudobolometric light curve for SN 2013ge assuming $E(B-V)_{\rm tot} = 0.067$\,mag. Although not the faintest stripped-envelope explosion ever seen, the bolometric light curve of SN 2013ge is fainter by nearly an order of magnitude compared to many other SNe~Ib/c. It also evolves more slowly and shows two distinct components in the $u$ and UV bands, including a relatively long rise time ($\sim 4$--6\,days) for the first component.  To describe these characteristics, \citet{drout16} proposed different potential progenitor scenarios, including an extended envelope, a small ejection $< 1$\,yr prior to explosion, or an asymmetric ejection of a small amount of nickel-rich material at high velocities.

Here we present deep, late-time {\it HST}/WFC3 observations of the position of SN 2013ge more than 4 years after we expect the SN radioactive component to have faded, with an intent to search for a surviving companion. Section \ref{sec:observations} presents our {\it HST} observations and detection of a point source at the position of SN 2013ge. We include estimates for reddening, distance, and metallicity. Our analysis is given in Section \ref{sec:analysis}, where we fit the detected point source with an SED and consider various physical origins, including the progenitor star's surviving companion. Section \ref{sec:con} discusses our conclusions.

\section{Observations}
\label{sec:observations}


\begin{deluxetable*}{ l l l l ll l}
\caption{{\it HST}/WFC3 Imaging \label{tab:tab1}}
\tablehead{
\colhead{UT Date} & \colhead{MJD} & \colhead{Epoch} & \colhead{Filter} & \colhead{Exposure} & \colhead{SN Magnitude} & \colhead{Star B Magnitude}\\
\colhead{}     & \colhead{} & \colhead{(days)} &    \colhead{}   & \colhead{(s)}      & \colhead{Vega (err)} & \colhead{Vega (err)}
    }
\startdata
\hline
\multirow{2}{*}{20161031} & \multirow{2}{*}{57692} & \multirow{2}{*}{1088} & F814W & 780 & 23.84 (0.09) & 25.67 (0.25)\\
    &   &   & F555W & 710 & 24.41 (0.05)  & 25.54 (0.08)\\
\hline
\multirow{2}{*}{20170601} & \multirow{2}{*}{57905} & \multirow{2}{*}{1301} & F300X & 1200 & ---  & ---\\
    &   &   & F475X & 350 & ---  & ---\\
\hline
\multirow{2}{*}{20190502} & \multirow{2}{*}{58605} & \multirow{2}{*}{2001} & F814W & 780 & 24.79 (0.14)  & 25.77 (0.30)\\
    &   &   & F555W & 710 & 25.36 (0.11)  & 25.44 (0.12)\\
\hline
\multirow{4}{*}{20201026} & \multirow{4}{*}{59148} & \multirow{4}{*}{2544} & F275W & 13803 & 25.03 (0.13)  & 25.04 (0.09)\\
    &   &   & F336W & 8090 &  24.96 (0.05)  & 24.75 (0.05)\\
    &   &   & F438W & 1000 & 25.83 (0.15)  & 25.87 (0.17)\\
    &   &   & F555W & 700 & 26.23 (0.17)  & 25.57 (0.11)\\
\hline
\enddata
\end{deluxetable*}

Table \ref{tab:tab1} summarizes the observations obtained of SN 2013ge with the {\it HST} WFC3/UVIS channel as part of program GO-16165 (PI O. Fox), as well as archival data for GO-15166 (PI A. Filippenko), GO-14762 (PI J. Maund), and GO-14668 (PI A. Filippenko).  The individual UVIS {\tt flc} frames in all bands are obtained from the Barbara A. Mikulski Archive for Space Telescopes (MAST) at \url{http://dx.doi.org/10.17909/c55c-rc43}. The data followed standard pipeline processing. The frames in each band then have cosmic-ray hits masked and are combined into mosaics by running them through AstroDrizzle in PyRAF.

\begin{figure*}
\begin{center}
\includegraphics[width=6in]{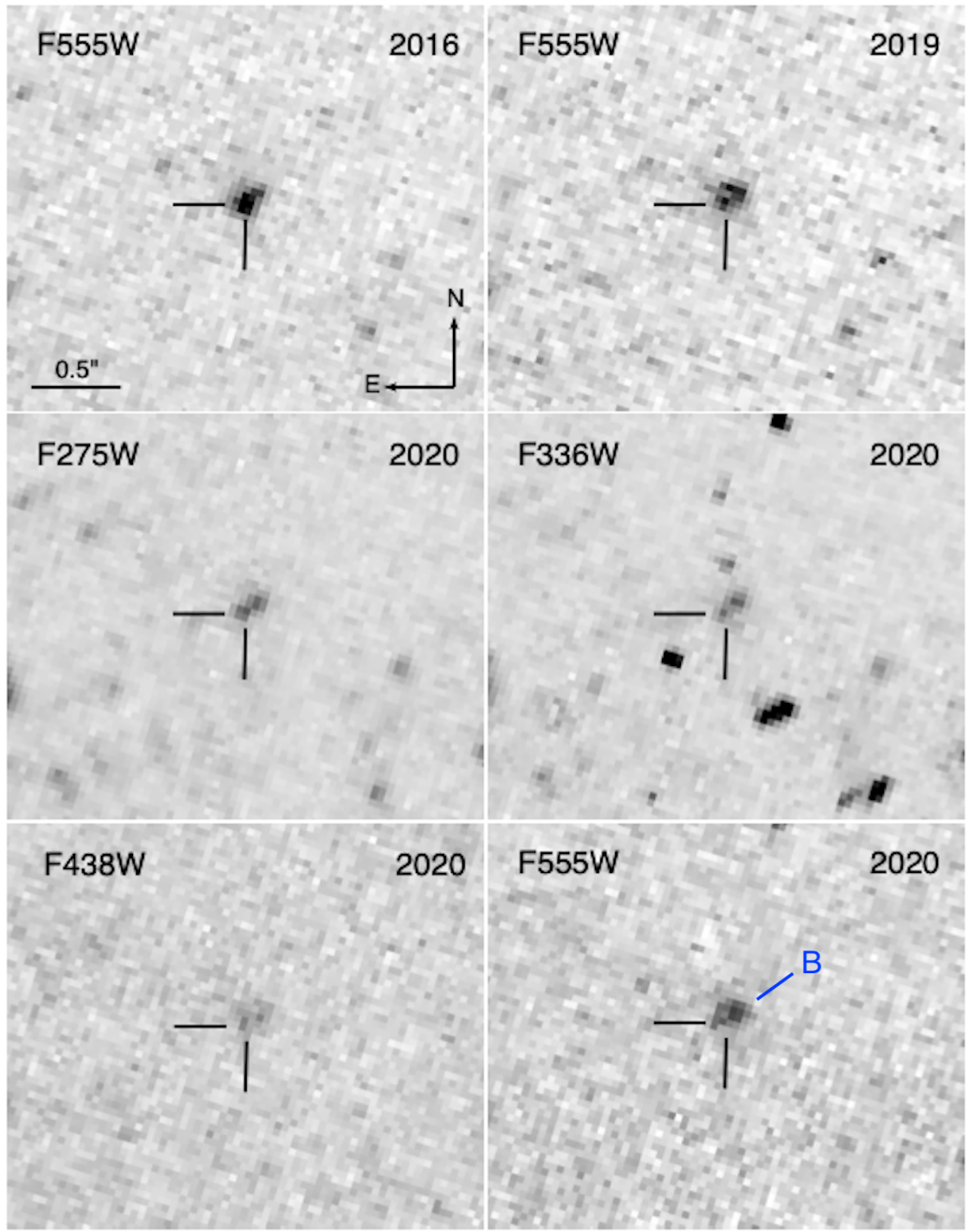}
\caption{\footnotesize Late-time {\it HST}/WFC3 UVIS imaging of the position of SN 2013ge spanning 2016 through 2020 as summarized in Table \ref{tab:tab1}. The F555W images offers the only constant filter throughout all three epochs and show the fading SN to the southeast of a constant, brighter source (``Source B''). We use the centroid of this fading source to obtain PSF photometry for the analysis in the paper, as described in the text. \label{fig:dimming}}
\end{center}
\end{figure*}

We locate the position of the SN in the new image mosaics using the {\it HST}/WFC3 data from 2016 and 2019 to track the SN as it fades in the F555W filter. Figure \ref{fig:dimming} identifies the fading source and a bright, neighboring source that does not get fainter, which we call ``Source B.'' The nature of Source B is not clear, but it is located 2.8 UVIS pixels from the fading source.

\subsection{{\it HST}/WFC3 Photometery}

We perform photometry by running DOLPHOT \citep{dolphin00} on individual ({\tt flc}) images, which are calibrated and aligned to a reference drizzled frame. Source detection is performed on a virtual stack, and photometry is obtained by forced point-spread-function (PSF) fitting to each source on the individual frames. The individual {\tt flc} measurements are merged to provide single measurements for each star in each observed bandpass. We make no charge-transfer efficiency (CTE) corrections as images have already been corrected for CTE prior to the photometry. We apply standard aperture corrections measured from the isolated bright stars in each exposure to account for any differences between the true PSF and the model. We did not measure photometry on the 2017 F300X and F475X data (GO-14762) because the combination of the ``X'' filters and subarray readout are not supported currently by DOLPHOT. The photometry for these filters is beyond the scope of this paper, but we list the data-set in Table \ref{tab:tab1} as a reference. 

\begin{figure}
\begin{center}
\includegraphics[width=3.5in]{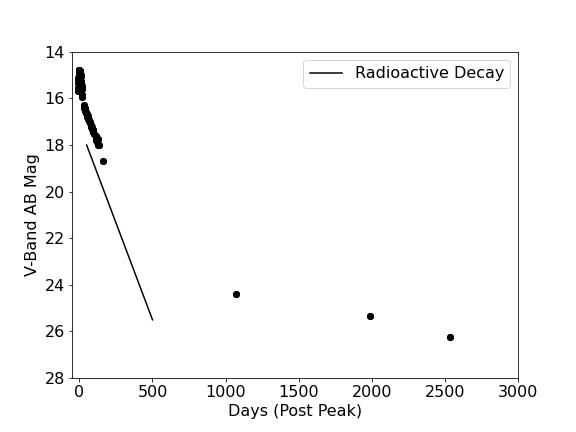}
\caption{\footnotesize $V$-band light-curve evolution of SN 2013ge. The early-time data are from \cite{drout16}, while the late-time photometry points were obtained by {\it HST} and listed in Table \ref{tab:tab1}. Overplotted is the typical slope from the radioactive decay \citep{dessart16}, scaled to the observed light-curve as a reference. The light curve shows signs of continued fading, which may signify some component of CSM shock interaction at these wavelengths. However, like for SN 1993J, the dominant blue component of the SED in Figure \ref{fig:sed} is evidence for the emerging component of a hot, blue source not associated with CSM interaction that we consider to be the putative progenitor companion. This scenario is discussed further in Section \ref{sec:interaction}.}
\label{fig:lightcurve}
\end{center}
\end{figure}

Table \ref{tab:tab1} lists the photometry for both the fading source and Source B. Given that the F555W and F814W photometry of Source B is consistent with no change of flux, combined with the fact that the location of Source B is separated from the candidate companion by more than 2 pixels, we conclude that Source B is not significantly contributing to the photometry of the fading source. Figure \ref{fig:lightcurve} plots the F555W photometry of the fading source for future reference.

\subsection{Distance and Reddening Estimates}
\label{sec:reddening}

{\it Distance:} \citet{drout16} adopted a distance of $23.7 \pm 1.7$\,Mpc, corresponding to the NED distance after correcting for Virgo, Great Attractor, and Shapley Supercluster infall, and H$_0$ = 73\,km\,s$^{-1}$\,Mpc$^{-1}$.  The distance, however, is quite uncertain. Cosmicflows-3 \citep{tully16} yields a much smaller value of 14.6\,Mpc, whereas the distance from the Numerical Action Methods model \citep[NAM;][]{kourkchi20} is 21.8\,Mpc. For lack of any more precise estimator of the distance, and given that the NAM distance is consistent (within the errors) with the value chosen by \citet{drout16}, we adopt that latter distance here as well. This distance serves as a conservative estimate since a closer distance would only correspond to a less-massive star.

{\it Reddening:} \citet{drout16} adopted a total reddening to SN\,2013ge of $E(B-V)_{\rm tot} = 0.067$\,mag and $R_V=3.1$. This consists of a Galactic foreground of $E(B-V)_{\rm Gal} = 0.02$\,mag from \citet{schlafy11}, and a host-galaxy contribution of $E(B-V)_{\rm host} = 0.047$\,mag based on the empirical relationship between the equivalent width (EW) of Na~{\sc i}~D and dust \citep{poznanski12}. Given that quantitative relations between Na~{\sc i}~D EW and $E(B-V)$ may be highly uncertain \citep{phillips13}, we explored other possible reddening and extinction laws, but found that different values did not significantly change any of our conclusions. For the rest of this paper, we therefore assume the same values used by \citet{drout16} listed above.

\begin{figure}
\begin{center}
\vspace{0.28in}
\includegraphics[width=3.5in]{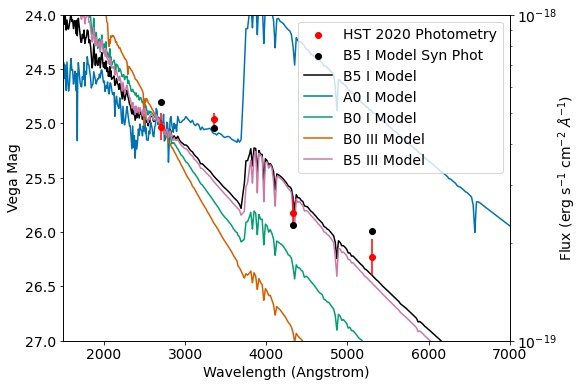}
\caption{\footnotesize Photometry of SN 2013ge obtained in 2020 (red circle; Table \ref{tab:tab1}). For comparison, overplotted are several different stellar models scaled to roughly the observed F275W value with an applied extinction $E(B-V)=0.067$\,mag and $R_V = 3.1$. Also shown are the synthetic photometry for the B5 I model (black circle). Specific stellar model properties are as follows (temperature, log $Z$~metallicity, and log $g$~gravity): B5 I (14000, -0.3, 2.44), A0 I (9730, -0.3, 2.14), B0 I (19000, -0.3, 3.0), B0 III (29000, -0.3, 3.5), B5 III (15000, -0.3, 3.5). The left and right ordinate axes are not directly aligned owing to the fact that any flux corresponds to a different Vega magnitude in each filter. The overplotted models should only be used for qualitative comparisons. While no model from the catalog matches perfectly to the observations, we find the SED is most consistent with a B5 I (supergiant) or III (giant).}
\label{fig:sed}
\end{center}
\end{figure}

\begin{figure*}
\begin{center}
\hspace{-0.6in}
\includegraphics[width=3.4in]{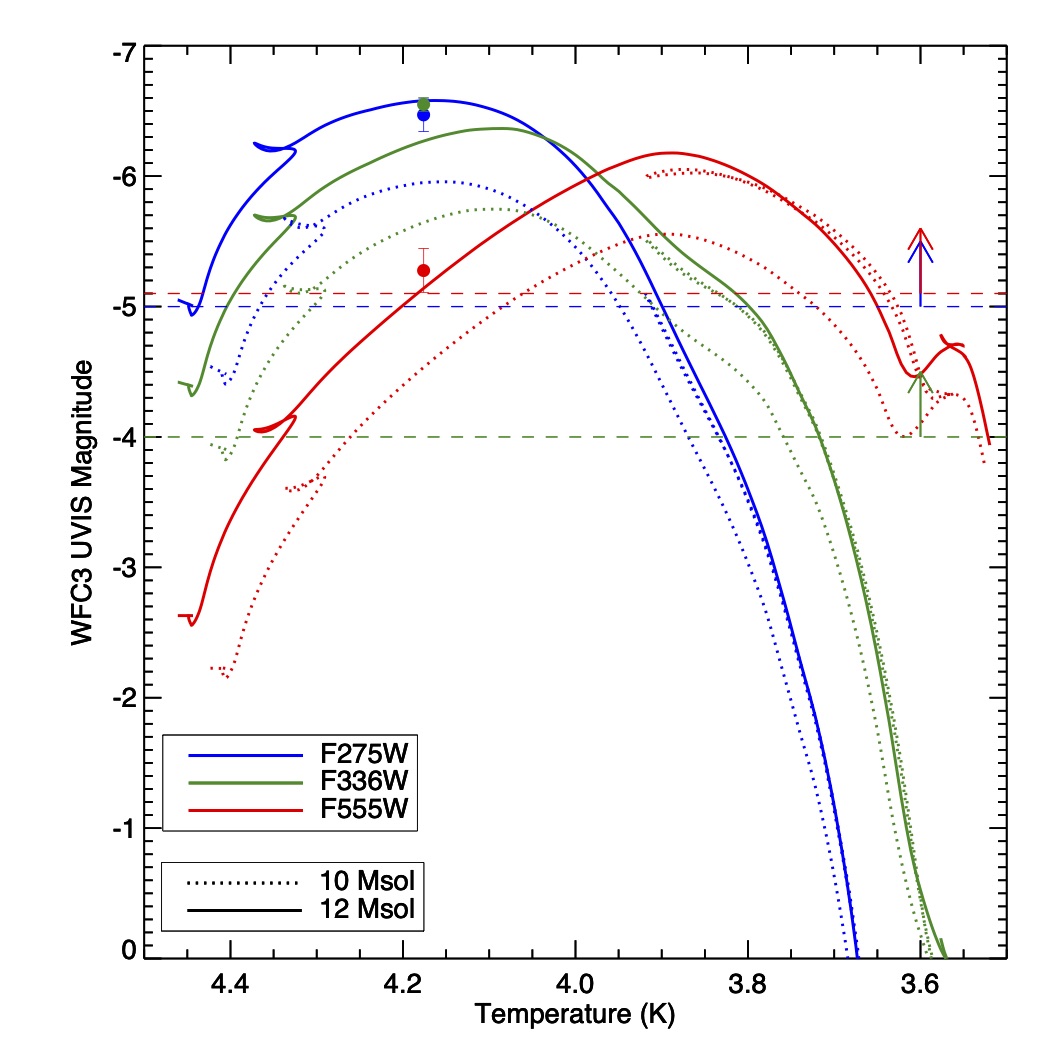}
\includegraphics[width=3.4in]{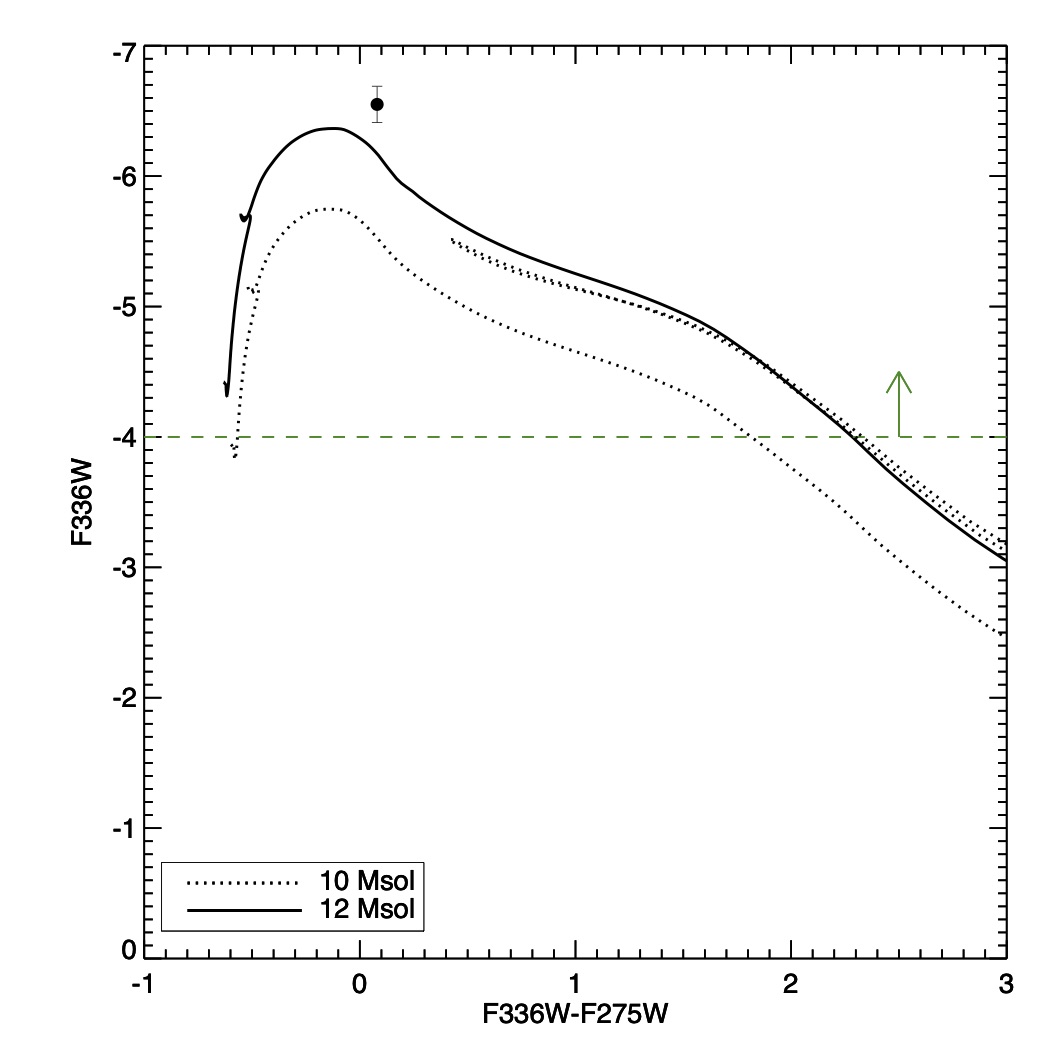}
\caption{\footnotesize Single-star evolutionary tracks of 10 and 12 \msolar\ stars in MIST for different {\it HST}/WFC3 filters, plotted in both ({\it left}) temperature and ({\it right}) color space. Overplotted are values for SN 2013ge (solid circles), in addition to the calculated limiting magnitudes of our designed observations (dashed horizontal lines). The source detected at the position of SN 2013ge is redward of the MS and consistent with $\sim 12$\,\msolar\ B5~I star crossing the Hertzsprung Gap (HG) toward becoming a giant, with an absolute magnitude F275W $= -6.5$ (Vega).}
\label{fig:cmd}
\end{center}
\end{figure*}

\section{Analysis}
\label{sec:analysis}

Figure \ref{fig:sed} plots the 2020 SED of the primary source in Figure \ref{fig:dimming}. We first consider the possibility that the source is the surviving companion to the progenitor of SN 2013ge, which would mark the first post-explosion direct detection of a surviving companion to a fully stripped-envelope SN~Ib/c. We also outline other possible scenarios, including a fading SN, CSM interaction, a line-of-sight coincidence, and a star cluster to which the progenitor belonged.

\subsection{The Companion Scenario}
\label{sec:companion}

We start with the initial assumption that the detected source is the surviving companion star of the primary star that exploded. We fit the photometry with spectra from Castelli and Kurucz Stellar Atmosphere Models \citep{castelli04} using {\tt stsynphot}\footnote{https://stsynphot.readthedocs.io/en/latest/index.html} and {\tt extinction}\footnote{https://extinction.readthedocs.io/en/latest/}. Figure \ref{fig:sed} shows comparisons of various stellar models. We find the best-fitting spectrum (i.e., lowest $\chi^2$) is a B5~I model ($T_{\rm eff} = 14,000$\,K and log\,g = 2.44), assuming a reddening consistent with an extinction of $E(B-V)\sim0.067$\,mag and $R_V=3.1$. We note that the synthetic photometry results do not perfectly align with the spectrum because a constant flux corresponds to a different Vega magnitude in each filter, so that the two ordinate axes (left and right) do not correspond one-to-one. The overplotted models should only be used for qualitative comparisons.

Figure \ref{fig:cmd} goes on to plot the absolute magnitudes of the source in a color-magnitude diagram (CMD). Overplotted as a reference are single-star evolutionary models for a 10 and 12\,\msolar\ star from MIST \citep{choi16}. Assuming that a single star currently dominates the light, the detected source is consistent with a 12\,\msolar\ B5~I-type star crossing the Hertzsprung Gap (HG) toward becoming a red supergiant, with an absolute magnitude F275W $= -6.5$ (Vega). We note that if the galaxy distance is indeed 14.6\,Mpc (see Section \ref{sec:reddening}), then the absolute magnitude would be dimmer by $\sim 0.7$\,mag, corresponding to a 10\,\msolar\ star.

The position on the CMD is redward of the MS. One possibility for this is that it is a very luminous MS star that is reddened by pre-existing or newly formed dust, although the optical spectra did not indicate significant reddening or evidence for late-time dust formation \citep{drout16}. Furthermore, a large mid-infrared sample of SNe indicates that dust formation is not common in SNe~Ib/c, most likely owing to the high velocities of these SE ejecta \citep[e.g.,][]{szalai19}.

A cool and inflated non-MS companion is unusual in existing model predictions because the companion would need to have a similar initial mass as the primary \cite{claeys11}. In this case, the companion may have also evolved past its MS at the moment of the progenitor's explosion. The star passes through this position during the hydrogen-shell burning phase on a relatively short thermal timescale, making the likelihood of it being at this evolutionary state at the time of the primary explosion $\lesssim 1.0\%$ \citep{zapartas17b}. Alternatively, the companion might be a blue central helium burning star. This is a slower evolutionary phase and hence more likely and has been suggested for stars in binaries that have gained mass    \citep[e.g.][]{justham14}. It is interesting to note that modeling of the pre-explosion photometry of the Type Ib SN 2019yvr also suggests that the potential companion is not an MS star \citep{sun21}.

Another possibility is that the companion was temporarily brought out of thermal equilibrium by the primary star, either pre-SN or post-SN. The thermal timescale is of the order of 10,000 years. 
In the pre-SN scenario, mass transfer from the primary would cause the companion to grow in size and become more luminous \citep[e.g.,][]{claeys11}. This scenario requires fine-tuning as the SN needs to have happened within a thermal timescale of the the last mass transfer event. In the post-SN scenario, interaction of the SN ejecta with the companion would leave the companion's envelope in an inflated state that gives it a cool and inflated position on the CMD \citep[e.g.,][]{ogata21}. 

\subsection{Fading Source and Shock Interaction}
\label{sec:interaction}

One common explanation for a late-time detection of a SN is that the source is still fading or is driven by shock interaction with a dense CSM, but such scenarios are typically associated with SNe with known shock interaction, such as SNe~IIn \citep[e.g.,][]{fox20}. The early spectra of SN 2013ge show no signs of CSM interaction, although one explanation for the unusual double peak may have been a small ejection $<1$\,yr prior to the explosion \citep{drout16}. While SNe~Ib/c typically explode in a low-density environment formed by years of progenitor winds, some SNe~Ib/c are known to exhibit early-time CSM interaction \citep[e.g., ][]{hosseinzadeh17}, and a growing number of late-time observations have revealed that some SNe~Ib/c may encounter a pre-existing shell years after the initial explosion \citep[e.g.,][]{milisavljevic12}, including SN 2019yvr discussed above \citep{sun21}. 

While SN 2013ge is too faint for late-time spectra, Figure \ref{fig:lightcurve} shows the light curve of SN 2013ge. The only consistent filter across all the late-time {\it HST} photometry is F555W (Table \ref{tab:tab1}). The fluxes are significantly above any fading emission from the radioactive decay, but the F555W photometry alone cannot rule out the possibility of declining shock interaction. 

The UV photometry offers a powerful tool for disentangling this scenario and is reminiscent of the Type IIb SN 1993J, which showed both a declining shock-interaction component at optical wavelengths and an emerging UV component from the surviving companion \citep{fox14}. The shape of the late-time SED for SN 2013ge in Figure \ref{fig:sed} is too blue to be explained by shock interaction alone, which is a flatter spectrum across all bands (see Fig.~11 of \citet{fox14}). While we cannot definitively rule out a shock-interaction component in the optical bands, any such component is not strongly impacting our UV analysis. 

\subsection{Line-of-Sight Coincidence or Stellar Cluster }
\label{sec:cluster}

The photometric SED of the putative stellar companion may not be easily distinguishable from some clusters or even just a handful of stars within a cluster. The WFC3 scale of 0\arcsec.0396\,pixel$^{-1}$ corresponds to 
$\sim 3.8$\,pc\,pixel$^{-1}$ at 20\,Mpc.  We consider several examples.  The association Ru~141 has a physical radius of $\sim 7$\,pc and a single dominant star \citep{camargo09}. If SN 2013ge exploded in such an environment, we estimate a $\sim 30$\% chance of having the unassociated star land on the same pixel as SN 2013ge. NGC 2004 (20\,Myr and $R_{\rm eff} \approx 6$\,pc) is one of the worst-case scenarios with $\sim 10$ stars brighter than $M_{\rm F275W} > -5$\,mag  \citep{niederhofer15}.  At a distance of $\sim 20$\,Mpc, such a cluster would be spread over $\sim 4$ pixels, implying the potential for up to 3 bright stars per pixel.  The radii of both of these clusters, however, are quite a bit smaller than the size of a typical OB association, which is often spread over $\sim 10$--20\,pc \citep{smith10carina}. Accounting for a factor of $\sim 3$--10 times larger in area, we therefore conclude that the likelihood of a chance alignment with an unresolved cluster or a bright star in a host cluster for our system is $<10\%$.

We can infer a minimum age of the progenitor system by assuming that all of the observed light is attributable to the most massive star in the system or the associated unresolved cluster. We assume our measured F438W luminosity of $-5.5$ is the maximum luminosity of any surviving companion (i.e., higher-mass stars would have yielded a larger luminosity). By comparing this luminosity to the PARSEC models \citep{bressan12}, it corresponds to an evolved star of 12\,\msolar\ and a cluster age of $\sim 20$\,Myr. This is consistent with our model fitting in Figures \ref{fig:sed} and \ref{fig:cmd}. If the flux is contributed by several stars, the cluster would be even older.

\section{Conclusion and Future Work}\label{sec:con}

This study reports the first potential post-explosion direct detection of a surviving companion star to a stripped-envelope SN of Type Ic (or possibly Ib). If the source is a single star, it is consistent with a $\sim 12$\,\msolar\ B5~I star with an absolute magnitude of F275W $= -6.5$ (Vega). The case for the surviving companion still requires additional observations and theoretical advancements. Similar to the case of SN~1993J \citep{fox14}, ongoing deep {\it HST} UV photometry will be able to rule out shock interaction by confirming that, even if the source fades in the optical, it is not fading in the UV. Similar to the analysis of SN~2017ein \citep{dyk18}, a complete SED of SN 2013ge should be compared to existing photometry of star clusters to rule out the possibility of any confusion from such a source. The prospect of further constraining the distance or reddening is slim given that {\it HST} cannot obtain observations of resolved stellar populations at this distance (e.g., the TRGB method), although such an experiment may be possible with {\it JWST}. From the theoretical perspective, the possibility of a cool, inflated non-MS companion must be better understood using existing binary evolution models, including BPASS \citep{eldridge17} and population synthesis \citep{zapartas17b}. These models, however, are limited by not accounting for accretion and generally not predicting evolved companions. In this case, a more empirical comparison to evolved stars with UV spectra is likely necessary.

Looking forward, the field would benefit from working on a larger sample. A comprehensive, statistically complete sample of direct companion observations is necessary to measure the fraction of surviving companions in binaries (which we refer to as the ``binary fraction''), the stellar types, and the mass distribution.  Deep upper limits also help in this respect.  Many of the  model variants of \citet{zapartas17b} predict a surviving companion in a majority of cases, and the associated mass distributions tend to be quite broad, peaking at $\sim 10$--12\,\msolar.  Deep upper limits down to the detection threshold at this particular mass range present a testable hypothesis: observations of historical stripped-envelope SNe should result in a nondetection roughly half the time (assuming a roughly symmetric distribution).  As the sample grows, the binary fraction can more thoroughly constrain the various models, and ultimately distinguish between stellar progenitor and binary evolution scenarios.
 
\section*{Acknowledgements}

This research is based on observations made with the NASA/ESA {\it Hubble Space Telescope} obtained from the Space Telescope Science Institute (STScI), which is operated by the Association of Universities for Research in Astronomy, Inc., under NASA contract NAS 5–26555. These observations are associated with programs GO-16165 (PI O. Fox), GO-15166 (PI A. Filippenko), GO-14762 (PI J. Maund), and GO-14668 (PI A. Filippenko); we are also grateful for the associated funding from NASA/STScI.  O.D.F. would like to thank Karl Gordon for useful conversations on dust. E.Z. acknowledges support from the Swiss National Science Foundation Professorship grant (project PP00P2 176868; PI Tassos Fragos). M.R.D. acknowledges support from the NSERC through grant RGPIN-2019-06186, the Canada Research Chairs Program, the Canadian Institute for Advanced Research (CIFAR), and the Dunlap Institute at the University of Toronto. In addition to the NASA/STScI funding, A.V.F. received support from the Miller Institute for Basic Research in Science (where he was a Miller Senior Fellow) and the Christopher R. Redlich Fund. D.~M.\ acknowledges NSF support from grants PHY-1914448 and AST-2037297. 

\bibliography{references.bib}{}
\bibliographystyle{aasjournal}

\end{document}